\documentclass[aps,twocolumn,showpacs,prl,amsmath,amssymb,floatfix,superscriptaddress]{revtex4-2}
\usepackage{amsmath,amssymb,graphicx,color}
\usepackage{bm}
\usepackage{bbm}
\usepackage[caption=false]{subfig}
\usepackage[usenames,dvipsnames]{xcolor}
\usepackage[colorlinks=true,citecolor=Cerulean,linkcolor=RubineRed,urlcolor=Cerulean]{hyperref}
\usepackage[normalem]{ulem}
\usepackage{soul,xcolor}
\usepackage{multirow}

\usepackage{float}

\newcommand{\Tr}{\ensuremath{\mathrm{Tr}\,}}

\newcommand{\ket}[1]{\ensuremath{\vert #1 \rangle}}
\newcommand{\hrho}{\ensuremath{\hat{\rho}}}

\newcommand{\bra}[1]{\ensuremath{\langle #1 \vert}}

\usepackage{fixltx2e,amsmath}
\MakeRobust{\eqref}

\usepackage{mathptmx} 

\DeclareMathAlphabet{\mathcal}{OMS}{cmsy}{m}{n}
\DeclareSymbolFont{largesymbols}{OMX}{cmex}{m}{n}

\begin{document}
\setstcolor{blue}
\title{Scaling law of asymptotic freedom in collective charging of quantum batteries}

\author{Gentaro Watanabe}
\affiliation{Department of Physics and Zhejiang Institute of Modern Physics, Zhejiang University, Hangzhou, Zhejiang 310027, China}
\affiliation{Zhejiang Province Key Laboratory of Quantum Technology and Device, Zhejiang University, Hangzhou, Zhejiang 310027, China}

\author{Chunlin Chen}
\affiliation{Department of Physics and Zhejiang Institute of Modern Physics, Zhejiang University, Hangzhou, Zhejiang 310027, China}

\author{B. Prasanna Venkatesh}
\affiliation{Indian Institute of Technology Gandhinagar, Palaj, Gujarat 382055, India}

\date{\today}

\begin{abstract}
We establish a universal scaling law for collective charging of quantum batteries, independent of microscopic details. We prove that the ergotropy-to-energy ratio approaches unity at least as fast as $\sim N^{-1}$ with the number of batteries $N$, implying generic asymptotic freedom. We further show how the universal $1/N$ scaling can be overcome: when the battery state becomes asymptotically pure, the convergence can be substantially faster, including $\sim N^{-b}$ with $b>1$ and even exponential scaling in $N^2$. Rigorous finite-$N$ upper and lower bounds on the ergotropy-to-energy ratio are further derived, providing nonasymptotic guarantees for the universal $1/N$ scaling.
\end{abstract}

\maketitle

\textit{Introduction}---Quantum batteries are energy-storage devices composed of quantum systems \cite{Campaioli2018,Bhattacharjee2021,Myers2022,Campaioli2024,Quach2023,Ferraro2026}. Over the past decade, they have become an active area of research in quantum thermodynamics, with studies exploring how quantum coherence and entanglement enhance energy storage and extraction \cite{Allahverdyan2004, Alicki2013, Hovhannisyan2013, Francica2017, Andolina2019, Andolina2019classical, GarciaPintos2020, Kamin2020, Francica2020, Wang2025, Shi2022, Salvia2023, Castellano2024, Gyhm2024}, and how environmental interactions can be engineered into a resource rather than treated merely as a source of decoherence \cite{Farina2019, Barra2019, Pirmoradian2019, Liu2019, Quach2020, Tabesh2020, Kamin2020b, Hovhannisyan2020, Carrega2020, Caravelli2021, Gherardini2020, Ghosh2021, Mitchison2021, Xu2021, Morrone2023, Morrone2023daemonic, Centrone2023, Rodriguez2024, Ahmadi2024, Cavaliere2025, Song2025, Shastri2025}. At the same time, experimental implementations of quantum 
battery protocols have been realised across a wide variety 
of platforms, including organic microcavities \cite{Quach2022, Tibben2025, Hymas2026}, superconducting qubits \cite{Hu2022, Ge2023,Li2025SCqubits,Gemme2022, Gemme2024, Niu2024, Razzoli2025, Elyasi2025}, NMR \cite{Joshi2022, Cruz2022}, and quantum dots \cite{Wenniger2023}.

Practical quantum batteries must store macroscopic amounts of energy, making scalable many-body architectures essential: a single microscopic quantum system holds only a negligible amount of energy and power. This led to a line of work on collectively charged multipartite batteries \cite{Binder2015,Campaioli2017,Ferraro2018,LeSpinChain2018,Julia-Farre2020,Rossini2020,Gyhm2022,Hokkyo2024,Watanabe2020,Peng2021,Gao2022,Mayo2022,Ueki2022,Zhang2023,Zhang2024,Li2025,Ito2020,Yang2024,Pokhrel2025,Canzio2025,Wang2025,Purkait2026}, beginning with the demonstration that global charging operations can yield superextensive charging power scaling with the number of cells $N$ \cite{Binder2015, Campaioli2017, Ferraro2018, LeSpinChain2018}. Collective charging has been demonstrated in cavity-QED platforms \cite{Quach2022, Hymas2026}, and theoretical work has established that a genuine quantum advantage beyond collective classical effects requires non-local many-body interactions, as exemplified by the Sachdev--Ye--Kitaev battery \cite{Julia-Farre2020, Rossini2020,Gyhm2022,Hokkyo2024}. These developments motivate the search for universal large-$N$ properties of collectively charged quantum batteries.

\begin{figure}[t!]
\centering
\includegraphics[width=0.99 \columnwidth]{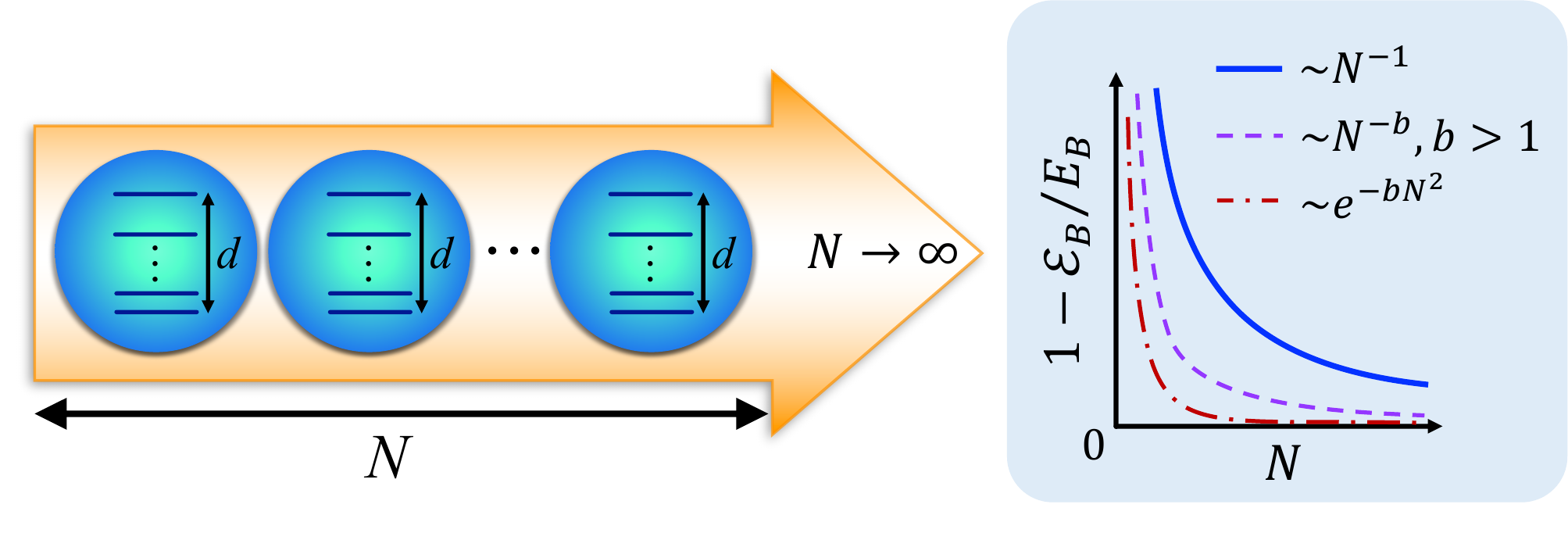}
\caption{
Emergence of asymptotic freedom in collective quantum battery charging. Left: an ensemble of $N$ identical $d$-level batteries with arbitrary spectra. Right: representative scaling of the ergotropy-to-energy ratio $\mathcal{E}_B/E_B$ toward unity with increasing $N$, including the generic $\sim N^{-1}$ scaling and accelerated convergence.
}
\label{fig:setup}
\end{figure}

An important large-$N$ question concerns the extractability of stored energy 
in collectively charged quantum batteries. Remarkably, in certain models the 
ratio of ergotropy to stored energy, $\mathcal{E}_B/E_B$, 
approaches unity as $N\to\infty$, implying that nearly all stored energy becomes 
extractable as work \cite{Andolina2019}. 
This phenomenon, known as \emph{asymptotic freedom}, points to a
universal advantage of collective quantum batteries as scalable energy-storage 
devices. However, these studies addressed specific models, leaving 
the question of universality of asymptotic freedom open 
\cite{Andolina2019,Ito2020,Purkait2026,Wang2025,Yang2024,Pokhrel2025,Canzio2025}. In this Letter, we establish that asymptotic freedom is generic 
for broad classes of collectively charged quantum batteries schematically illustrated 
in Fig.~\ref{fig:setup}: the deficit $1 - \mathcal{E}_B/E_B$ 
vanishes at least as fast as $\sim N^{-1}$. We further identify asymptotic purity of the  battery state as the key condition enabling faster convergence beyond the generic $1/N$ scaling; we also 
derive rigorous finite-$N$ upper and lower bounds on 
$\mathcal{E}_B/E_B$.

\textit{Setup}---We consider collective charging of $N$ identical quantum batteries (QBs) as depicted in Fig.~\ref{fig:setup}, each a copy of the same system with no mutual interactions. Because all $N$ copies are identical and charged in the same manner, the dynamics and the resulting state are permutation invariant with respect to the batteries. Throughout this work, we assume that the Hilbert space of each battery is finite, with dimension $d$, and that the battery has a nondegenerate ground state. Within this setting, the spectral decomposition of the total Hamiltonian $\hat{H}_B$ and density operator $\hat{\rho}_B$ of the $N$ QBs can be written in general as
\begin{align}
  \hat{H}_B = \sum_{i \ge 0} \epsilon_i^\uparrow |\epsilon_i^\uparrow\rangle \langle \epsilon_i^\uparrow|\,,\quad 
  \hat{\rho}_B = \sum_{i \ge 0} \eta_i^\downarrow |\eta_i^\downarrow\rangle \langle \eta_i^\downarrow|\,,
\end{align}
where the eigenvalues are ordered as $\epsilon_i^\uparrow \le \epsilon_{i+1}^\uparrow$ and $\eta_i^\downarrow \ge \eta_{i+1}^\downarrow$. Without loss of generality, we set the ground-state energy to zero: $\epsilon_0^\uparrow = 0$, and denote the energy gap between the ground and first excited states by $\Delta\epsilon_0 \equiv \epsilon_1^\uparrow - \epsilon_0^\uparrow = \epsilon_1^\uparrow > 0$.

\textit{Energy \& Ergotropy}---The total energy $E_B$ of the $N$-QB system is given by
\begin{equation}
  E_B \equiv \Tr[\hat{H}_B\, \hat{\rho}_B]\,.
\end{equation}
Since there is no interaction between the batteries and the state $\hat{\rho}_B$ is permutation invariant under exchange of the batteries, $E_B$ is extensive,
\begin{equation}
  E_B = N e_B \propto N\,,
\end{equation}
where $e_B$ denotes the energy of a single battery. A proof of extensivity and the explicit definition of $e_B$ are provided in Appendix~\ref{app:extensivity}. The ergotropy $\mathcal{E}_B$ of the $N$-QB system is defined as the maximum amount of work that can be extracted via a cyclic unitary operation~\cite{Allahverdyan2004},
\begin{equation}
  \mathcal{E}_B \equiv E_B - \min_{\hat{U}} \Tr[\hat{H}_B\, \hat{U}\,\hat{\rho}_B\,\hat{U}^\dagger] = E_B - \Tr[\hat{H}_B\, \hat{\rho}_B^\downarrow]\,,
\end{equation}
where the minimization $\min_{\hat{U}}$ is taken over all unitary operators acting on the $N$-QB system and $\hat{\rho}_B^\downarrow$ denotes the passive state associated with $\hat{\rho}_B$ and is given by 
\begin{align}
    \hrho_B^\downarrow = \sum_{i \ge 0} \eta_i^\downarrow \ket{\epsilon_i^{\uparrow}}\bra{\epsilon_i^{\uparrow}}.
\end{align}
The figure of merit we focus on is the ratio $\mathcal{E}_B/E_B$, whose deviation from unity signifies the amount of locked or inaccessible energy of the QBs.

Whether asymptotic freedom is generic or model-dependent is a key open question, which we settle with the following universal scaling theorem:\\
\textit{Theorem 1}---For $N$ copies of a QB, each consisting of an arbitrary finite-dimensional quantum system of dimension $d$, the ratio $\mathcal{E}_B/E_B$ asymptotically approaches unity as $N$ increases as
\begin{equation}
  \frac{\mathcal{E}_B}{E_B} = 1 - \frac{a}{N} \,, \quad (N \rightarrow \infty) \,,\label{eq:theorem1}
\end{equation}
or faster, where $a > 0$ is an $N$-independent constant.\\
Although $\mathcal{E}_B/E_B$ generally depends on the microscopic details of the QBs, Theorem~1 shows that the leading $1/N$ scaling of $1-\mathcal{E}_B/E_B$ is universal. In what follows, we develop a proof for Theorem 1 by showing that both the upper and lower bounds of $1 - \mathcal{E}_B/E_B$ asymptotically scales as $\sim 1/N$ to leading order, provided that the battery state remains nonpure for arbitrary $N$.

\textit{Upper bound of $\mathcal{E}_B/E_B$}---As the first step towards the central result of this letter, we will derive an upper bound on the extractable fraction of the battery energy $\mathcal{E}_B/E_B$. As a preparation, we introduce the residual population of the $N$-battery system defined as
\begin{equation}
  \delta(N) \equiv 1 - \eta_0^\downarrow\,,
\end{equation}
which represents the total population in the excited states in the passive state $\hat{\rho}_B^\downarrow$. If $\delta(N)=0$ for some finite $N$, the battery state $\hat{\rho}_B$ is pure and the system is already free at that $N$ (i.e., $\mathcal{E}_B/E_B =1$). In this trivial case, the convergence is obviously faster than $1/N$. We therefore focus on the nontrivial situation in which $\delta(N) \ne 0$ for all finite $N$. 
We define
\begin{equation}
  \delta_{\mathrm{min}} \equiv \min_N \delta(N)\,.
\end{equation}
Thus, if $\delta_{\mathrm{min}} = 0$, the value zero is not attained at any finite $N$, but only approached asymptotically as $N \rightarrow \infty$.

The minimum passive energy is realized when all the residual population $\delta(N)$ is in the first excited state with $\epsilon_1^\uparrow$.
Using $\delta_{\mathrm{min}}$, the passive energy can be lower bounded as
\begin{align}
  E_B - \mathcal{E}_B &= \Tr\left[\hat{H}_B\, \hat{\rho}_B^\downarrow\right] = \sum_{i \ge 0} \epsilon_i^\uparrow\, \eta_i^\downarrow = \sum_{i\ge 1} \epsilon_i^\uparrow\, \eta_i^\downarrow \nonumber\\
  &\ge \epsilon_1^\uparrow\, \delta(N) \ge \epsilon_1^\uparrow\, \delta_{\mathrm{min}} = \Delta\epsilon_0\, \delta_{\mathrm{min}}\,.
\end{align}
From the first to the second line, we have used the assumption of the absence of ground-state degeneracy. Using the extensivity of $E_B$, i.e., $E_B = e_B N$, we obtain our first result upper-bounding $  \frac{\mathcal{E}_B}{E_B}$ (valid for any finite $N$):
\begin{equation}
  \frac{\mathcal{E}_B}{E_B} \le 1 - \frac{\Delta\epsilon_0}{e_B}\, \frac{\delta_{\mathrm{min}}}{N}\,.\label{eq:ratio_upperbound1}
\end{equation}

Focusing next on the limit $N\rightarrow \infty$, we define the asymptotic residual population
\begin{equation}
  \delta_\infty \equiv \lim_{N\rightarrow\infty}\delta(N)\,,
\end{equation}
when the limit exits. The classification below depends only on whether $\delta(N)$ vanishes asymptotically or not, and can therefore be made in terms of $\delta_\infty$: in what follows, we distinguish the cases $\delta_\infty \ne 0$ (Case 1) and $\delta_\infty = 0$ (Case 2). Situations where $\delta_\infty$ does not exist are dealt with in Appendix~\ref{app:no_delta_infty}.
\smallskip

\noindent\emph{\textbf{Case 1:} $\delta_\infty \ne 0$}\,---If $\delta(N)$ converges to a nonzero constant in the large-$N$ limit, the bound can be further tightened by $\delta_\infty$. In this case, the passive energy asymptotically satisfies
\begin{equation}
  \lim_{N\rightarrow\infty}\left(E_B - \mathcal{E}_B\right) = \lim_{N\rightarrow\infty} \sum_{i\ge1} \epsilon_i^\uparrow\, \eta_i^\downarrow \ge \Delta\epsilon_0\,  \delta_\infty\,,
\end{equation}
which yields the asymptotic upper bound:
\begin{equation}
  \frac{\mathcal{E}_B}{E_B} \le 1 - \frac{\Delta\epsilon_0}{e_B}\, \frac{\delta_{\infty}}{N}\,,\quad (N \rightarrow \infty)\,.\label{eq:ratio_upperbound2}
\end{equation}
Since the leading term of $\delta(N)$ is constant, any residual $N$-dependence of $\delta(N)$ only contributes higher-order corrections and is therefore negligible in the large-$N$ limit.

\smallskip
\noindent\emph{\textbf{Case 2:} $\delta_\infty = 0$}\,---Since $\delta(N)$ vanishes asymptotically, we write
\begin{equation}
  \delta(N) = \delta_\infty + \Delta\delta(N) = \Delta\delta(N)\,,\label{eq:nonzeropartdelta}
\end{equation}
where $\Delta\delta(N) > 0$ and $\lim_{N\rightarrow\infty} \Delta\delta(N) = +0$. The passive energy then asymptotically satisfies
\begin{equation}
  E_B - \mathcal{E}_B = \sum_{i\ge 1} \epsilon_i^\uparrow\, \eta_i^\downarrow \ge \Delta\epsilon_0\, \Delta\delta(N)\,,\quad
  (N\rightarrow\infty)\,.\nonumber
\end{equation}
Using the extensivity $E_B = e_B N$, we obtain
\begin{equation}
  \frac{\mathcal{E}_B}{E_B} \le 1 - \frac{\Delta\epsilon_0}{e_B}\, \frac{\Delta\delta(N)}{N}\,,\quad (N\rightarrow\infty)\,.\label{eq:ratio_upperbound3}
\end{equation}
Unlike Case 1, where $\delta_\infty > 0$, here $\delta_\infty = 0$ and $\Delta\delta(N)$ decreases asymptotically to zero. Consequently, for this case where $\delta(N)$ vanishes and hence the purity of the battery state approaches $1$ asymptotically, the upper bound of the ratio $\mathcal{E}_B/E_B$ approaches unity faster than $\sim 1/N$. Having demonstrated via Eqs.~\eqref{eq:ratio_upperbound1}, \eqref{eq:ratio_upperbound2}, and \eqref{eq:ratio_upperbound3} that the upper bound of $\mathcal{E}_B/E_B$ approaches unity as $\sim 1/N$ or faster, we next derive similar behavior for the lower bound.

\medskip
\textit{Lower bound of $\mathcal{E}_B/E_B$}---Due to the permutation symmetry among the batteries, the states of the system composed of $N$ identical copies of a $d$-level QB can be labeled by the set of populations $\{n_i\}$ of the individual levels, with $i=0$, $1$, $\cdots$, $d-1$, satisfying $n_0 + n_1 + \cdots + n_{d-1} = N$. Accordingly, the total number of distinct states of the permutation-invariant Hilbert space is
\begin{equation}
  D_d(N) = \binom{N+d-1}{d-1}\,.\label{eq:Dd}
\end{equation}
Thus, crucially we see that due to permutation symmetry, the QB's state is restricted to a smaller dimension of $D_d(N)$ than $d^N$.

The passive energy is maximized when the residual population $\delta(N)$ is uniformly distributed over the $D_d(N)-1$ excited states. This yields
\begin{align}
  E_B - \mathcal{E}_B \le \frac{\delta(N)}{D_d(N) - 1} \sum_{i=1}^{D_d - 1} \epsilon_i^\uparrow\,.\label{eq:epass_upperbound1}
\end{align}
To derive an upper bound on the sum $\sum_{i=1}^{D_d-1}\epsilon_i^\uparrow$ appearing on the right-hand side above, we first focus on qudits with equally spaced levels, for which the following lemma holds (proved in detail in Appendix~\ref{app:lemma1}).\medskip\\
\emph{Lemma 1}: For $N$ qudits with equally spaced levels (with level spacing denoted by $\omega_B$), the entire set of $D_d(N)$ states is completely and exactly exhausted by the states corresponding to the lowest $d$ energy eigenvalues of the $N$-copy system, $\bar{\epsilon}_0 = 0$, $\bar{\epsilon}_1 = \omega_B$, $\cdots$, $\bar{\epsilon}_{d-1} = (d-1)\omega_B$ (here $\bar{\epsilon}_i$ labels the distinct energy eigenvalues in increasing order, with degeneracies included).\medskip\\
Therefore, for $N$ qudits, we have
\begin{align}
  \sum_{i=1}^{D_d-1} \epsilon_i^\uparrow &= \sum_{j=0}^{d-1} j \omega_B N_{j}
  = \omega_B \frac{(N+d-1)!}{(N+1)\, (d-2)!\, (N-1)!}\, ,\label{eq:energysum_qudits}
\end{align}
where $N_j = D_{j+1}(N)-D_j(N) = \binom{N+j-1}{j}$ denotes the number of degenerate states associated with the $(j+1)$th energy
eigenvalue $\bar{\epsilon}_j = j \omega_B$. 

We next consider a general $d$-level QB with an arbitrary level structure, allowing for degeneracies except in the ground state.
Since there is no interaction between the batteries, each eigenenergy $\bar{\epsilon}_i$ of the $N$-copy system is given by a sum of the single-battery eigenenergies $\{\epsilon_j^{(1)}\}$. We define the ``excitation number'' $\nu$ of $\bar{\epsilon}_i$ as the sum of the corresponding indices $j$ of the single-battery eigenenergies.
For instance, for two copies of a three-level system with eigenenergies $\epsilon_j^{(1)}$ ($j=0$, $1$, $2$), the composite eigenenergies $\epsilon_0^{(1)}+\epsilon_1^{(1)}$, $\epsilon_1^{(1)}+\epsilon_1^{(1)}$, and $\epsilon_1^{(1)}+\epsilon_2^{(1)}$ (or $\epsilon_0^{(1)}+\epsilon_3^{(1)}$) have excitation numbers $\nu=1$, $2$, and $3$, respectively. While for equally spaced qudits all $D_d(N)$ states satisfy $\nu \le d-1$ as claimed by Lemma 1, for QBs with uneven level spacings states with $\nu > d-1$ may also appear among the lowest $D_d(N)$ levels. Consequently, the total energy of the $D_d(N)$ states with $\nu \le d-1$ provides an upper bound (see Appendix~\ref{app:energyspectrum} for details),
\begin{equation}
  \sum_{i=0}^{D_d-1}\epsilon_i^\uparrow \le \text{(total energy of states with $\nu \le d-1$)}\,.\label{eq:energysum1}
\end{equation}
Let $\Delta \epsilon_\mathrm{max}$ denote the largest level spacing of a single battery, so that any state with excitation number $\nu = j$ satisfies $\bar{\epsilon}_i(\nu = j) \le j \Delta\epsilon_\mathrm{max}$. Combining this bound with the fact that the number of states with $\nu = j$ equals $N_j$ (from Lemma 1), Eq.~(\ref{eq:energysum1}) directly implies
\begin{align}
  \sum_{i=0}^{D_d-1}\epsilon_i^\uparrow
  &\le \Delta\epsilon_{\mathrm{max}} \sum_{j=0}^{d-1} j\, N_j
  = \Delta\epsilon_{\mathrm{max}} \frac{(N+d-1)!}{(N+1)\, (d-2)!\, (N-1)!}\,.\label{eq:energysum2}
\end{align}
Substituting Eq.~\eqref{eq:energysum2} into Eq.~(\ref{eq:epass_upperbound1}) and using the inequality $\frac{N}{N+1} \frac{(N+d-1)!}{(N+d-1)! - (d-1)!\, N!} \le 1$ (with equality only for $d=2$), we find
\begin{align}
  E_B - \mathcal{E}_B &\le (d-1) \frac{N}{N+1} \frac{(N+d-1)!}{(N+d-1)! - (d-1)!\, N!} \delta(N)\, \Delta\epsilon_{\mathrm{max}}\nonumber\\
  &\le (d-1)\, \delta(N)\, \Delta\epsilon_{\mathrm{max}}\,.\label{eq:epass_upperbound2}
\end{align}
Using the extensivity of $E_B$, Eq.~(\ref{eq:epass_upperbound2}) yields the following lower bound valid for any finite $N$:
\begin{equation}
  \frac{\mathcal{E}_B}{E_B} \ge 1 - (d-1) \frac{\Delta\epsilon_{\mathrm{max}}}{e_B}\frac{\delta(N)}{N}\,.\label{eq:ratio_lowerbound}
\end{equation}

As before, we now obtain asymptotic lower bounds from Eq.~(\ref{eq:ratio_lowerbound}) by distinguishing the cases according to the value of $\delta_\infty$:

\smallskip
\noindent\emph{\textbf{Case 1:} $\delta_\infty \ne 0$}\,--- 
If $\delta(N)$ converges to a nonzero constant $\delta_\infty$ in the large-$N$ limit, the asymptotic bound reads
\begin{equation}
  \frac{\mathcal{E}_B}{E_B} > 1 - (d-1) \frac{\Delta\epsilon_{\mathrm{max}}}{e_B}\frac{\delta_\infty}{N}\,, \quad (N \rightarrow \infty)\,.\label{eq:ratio_lowerbound2}
\end{equation}
\smallskip
\noindent\emph{\textbf{Case 2:} $\delta_\infty = 0$}\,---
In this case, we have
\begin{equation}
  \frac{\mathcal{E}_B}{E_B} \ge 1 - (d-1)\frac{\Delta\epsilon_{\mathrm{max}}}{e_B}\frac{\Delta\delta(N)}{N}\,, \quad (N \rightarrow \infty)\,.\label{eq:ratio_lowerbound3}
\end{equation}
with $\Delta \delta(N)$ defined in Eq.~\eqref{eq:nonzeropartdelta}.

\textit{Proof of Theorem 1}---For $\delta_\infty \neq 0$, 
Eqs.~(\ref{eq:ratio_upperbound2}) and 
(\ref{eq:ratio_lowerbound2}) both scale as $1/N$, giving 
Eq.~(\ref{eq:theorem1}) with $\delta_\infty\,
\Delta\epsilon_0/e_B \le a < (d-1)\,\delta_\infty\,
\Delta\epsilon_{\mathrm{max}}/e_B$. For $\delta_\infty 
= 0$, Eqs.~(\ref{eq:ratio_upperbound3}) and 
(\ref{eq:ratio_lowerbound3}) give
\begin{equation}
  \frac{\mathcal{E}_B}{E_B} = 1 - 
  a'\frac{\Delta\delta(N)}{N}\,, \quad (N \rightarrow 
  \infty)\,,\label{eq:ratio_fastconv}
\end{equation}
with $\Delta\epsilon_0/e_B \le a' \le (d-1)\,
\Delta\epsilon_{\mathrm{max}}/e_B$, and since 
$\Delta\delta(N) \to 0$, the ratio approaches unity 
faster than $\sim 1/N$. \hfill $\blacksquare$

\bigskip

The asymptotic scaling is thus governed entirely by 
whether $\delta_\infty \neq 0$ or $\delta_\infty = 0$, 
which correspond directly to the battery state remaining 
mixed or becoming asymptotically pure, since 
$\mathrm{Tr}\hat{\rho}^2_B \to 1\,(\neq 1)$ when 
$\delta_\infty = 0\,(\neq 0)$. For the case 
$\delta_\infty \neq 0$, Eq.~(\ref{eq:theorem1}) yields 
the following corollary:

\noindent \textit{Corollary 1}---If the battery state $\hat{\rho}_B$ remains nonpure for all $N$ (including $N \rightarrow \infty$), the ratio $\mathcal{E}_B/E_B$ asymptotically approaches unity as per Eq.~(\ref{eq:theorem1}), with the bounds on $a$ given in the proof of Theorem~1.

We note that in the special case of $d = 2$, the lower bound is saturated: $a = (\Delta \epsilon_0 / e_B)\,\delta_\infty$, reflecting the fact that the number of distinct states $D_2(N)=N+1$ is exactly exhausted by the single ground state $\epsilon_0^\uparrow$ and the $N$ degenerate first-excited states $\epsilon_1^\uparrow$. This implies $\eta_i^\downarrow = 0$ for $i\ge 2$, so that $\delta(N)=\eta_1^\downarrow$.

\begin{figure}[t!]
\centering
\includegraphics[width=0.7 \columnwidth]{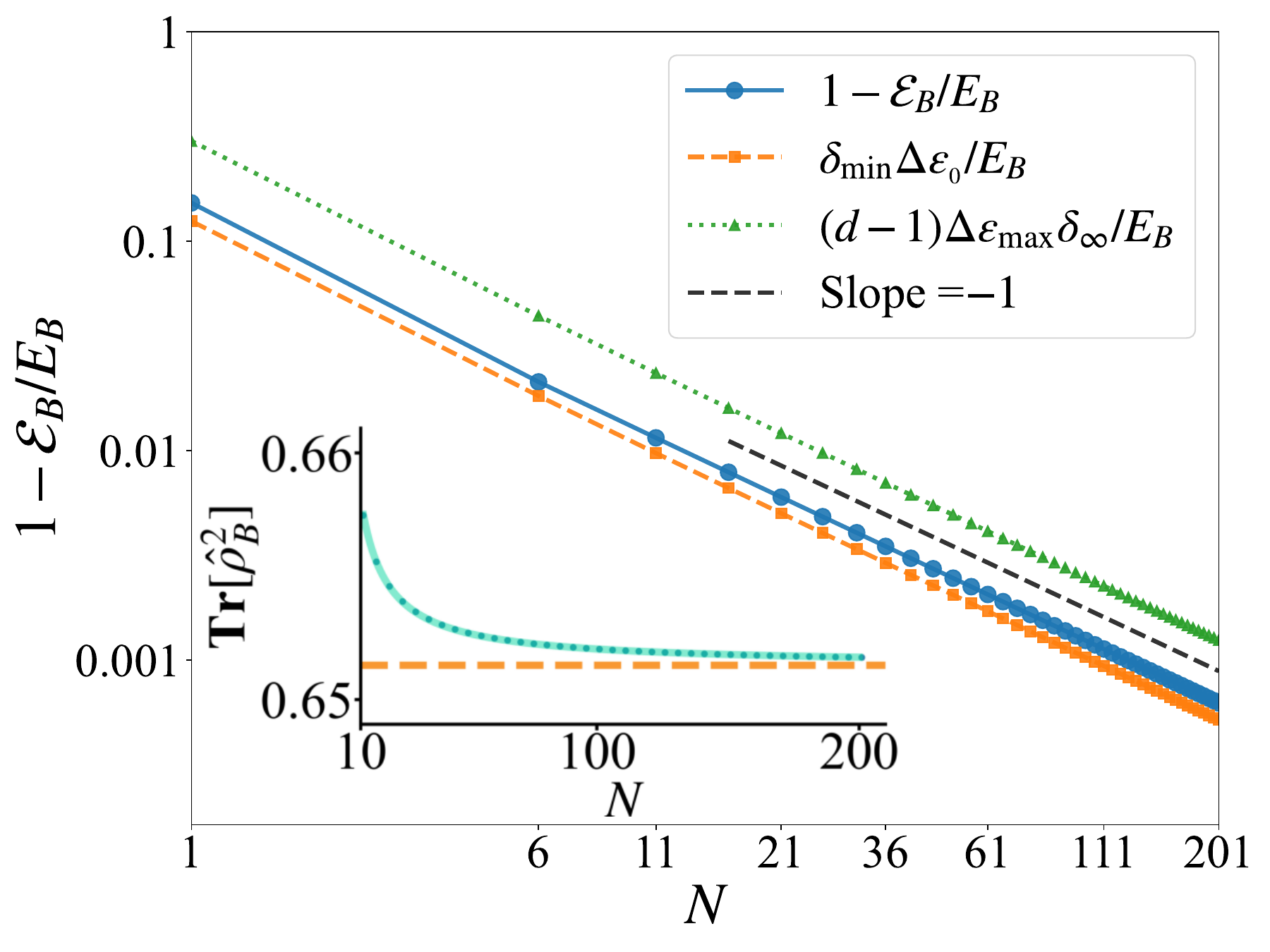}
\caption{Demonstration of $1/N$ scaling toward asymptotic freedom for $1-(\mathcal{E}_B/E_B)$, together with the bounds given by Eqs.~(\ref{eq:ratio_upperbound1}) and (\ref{eq:ratio_lowerbound2}), for unitary charging of the $d=3$ Dicke model. The inset shows the purity as a function of $N$. The parameter values are $\omega_C = \omega_B$ and $g=20\omega_B$.
}
\label{fig:1/N-scaling_and_bounds}
\end{figure}

\textit{Example 1: $1/N$ scaling toward asymptotic freedom}---As a concrete example to illustrate the $1/N$ scaling, we consider unitary charging of $N$ identical $d$-level spins in the $d$-level Dicke and Tavis--Cummings (TC) models, where a single bosonic cavity mode acts as a charger.
The total Hamiltonian $\hat{H} = \hat{H}_B + \hat{H}_C + \hat{H}_{BC}$ consists of the battery Hamiltonian $\hat{H}_B = \sum_{i=1}^N \hat{h}_{B_i}$, with $\hat{h}_{B_i} = \omega_B (\hat{s}_{z,\, i} + s)$ and $s = (d-1)/2$, the charger Hamiltonian $\hat{H}_C = \omega_C\, \hat{a}^\dagger \hat{a}$, and the battery-charger coupling Hamiltonian $\hat{H}_{BC}$. Here, $\hat{s}_{z,\,i}$ denotes the $z$-component of the spin-$s$ operator acting on the $i$th $d$-level spin, $\omega_B$ is the level spacing of the spins, $\omega_C$ is the cavity mode frequency, and $\hat{a}^\dagger$ ($\hat{a}$) is the creation (annihilation) operator of a cavity photon. We consider the resonant case with $\omega_B = \omega_C$. For the Dicke model, the coupling Hamiltonian is given by $\hat{H}_{BC} = \frac{g}{\sqrt{N}} (\hat{a}^\dagger + \hat{a}) (\hat{S}_+ + \hat{S}_-)$, while for the TC model, $\hat{H}_{BC} = \frac{g}{\sqrt{N}} (\hat{a}^\dagger \hat{S}_- + \hat{a} \hat{S}_+)$. Here, $g$ is the coupling strength, and $\hat{S}_\pm \equiv \sum_{i=1}^N \hat{s}_{\pm,\, i}$ and $\hat{S}_z \equiv \sum_{i=1}^N \hat{s}_{z,\, i}$ are the collective operators satisfying the SU(2) algebra, $[\hat{S}_z,\, \hat{S}_\pm] = \pm \hat{S}_\pm$ and $[\hat{S}_+,\, \hat{S}_-] = 2 \hat{S}_z$.

\begin{figure}[t!]
\centering
\includegraphics[width=0.7 \columnwidth]{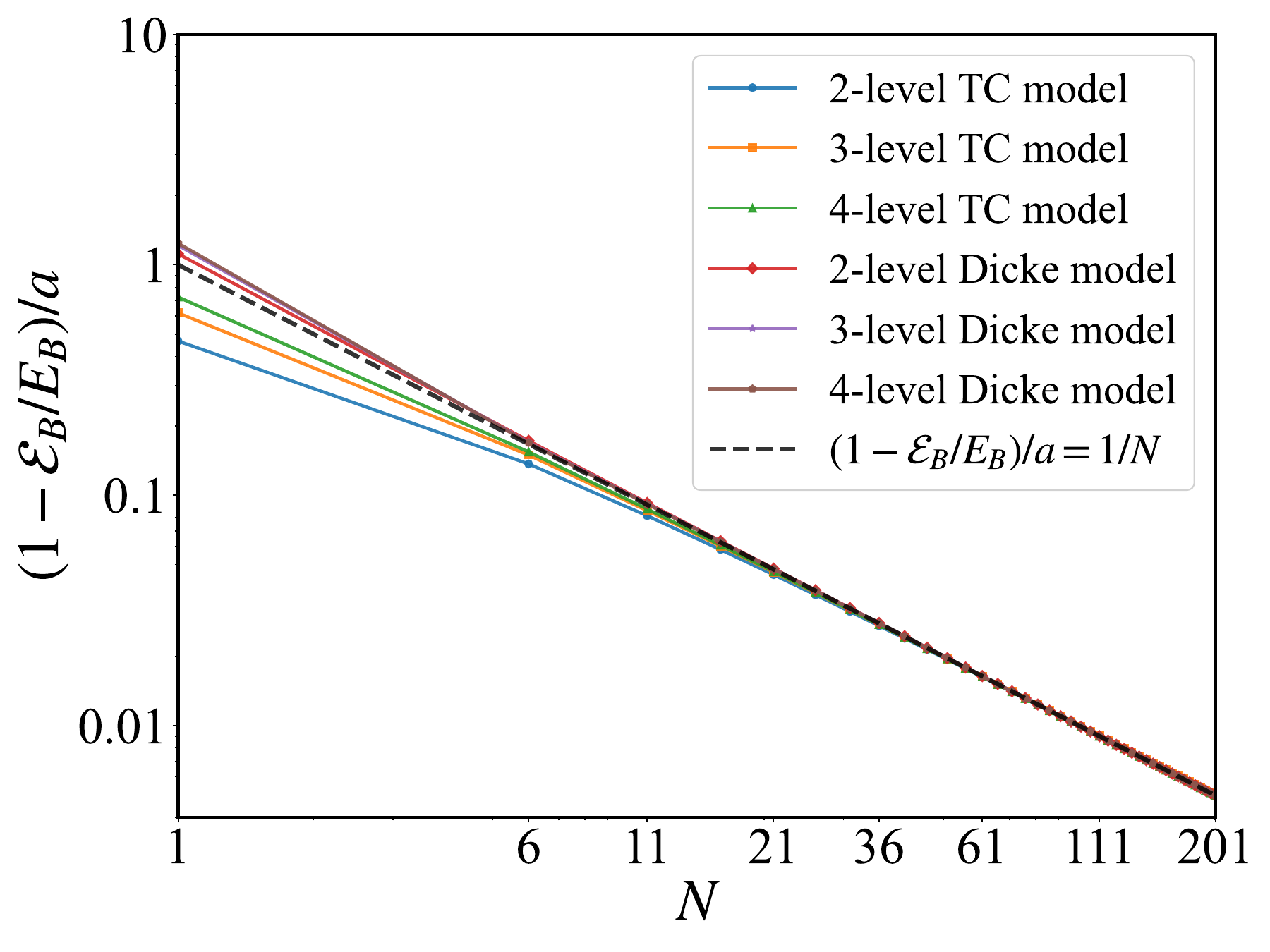}
\caption{Demonstration of the universal $1/N$ scaling toward asymptotic freedom. All models considered collapse onto a single line with slope $-1$ in the log-log plot. The parameter values are the same as Fig.~\ref{fig:1/N-scaling_and_bounds}.
}
\label{fig:collapse}
\end{figure}
We consider unitary charging from the ground state of all batteries with the cavity initialized in a coherent state containing $N(d-1)$ photons on average, corresponding to the maximum storable energy. We numerically solve the Schr\"odinger equation to evaluate the energy $E_B$ and ergotropy $\mathcal{E}_B$ at charging time $\tau$. Following Ref.~\cite{Andolina2019}, $\tau$ is defined as the time at which the charging power $P\equiv E_B(t)/t$ is maximized. We have verified that all the results remain qualitatively robust to the choice of $\tau$.

Figure~\ref{fig:1/N-scaling_and_bounds} shows the $N$-dependence of $1-(\mathcal{E}_B/E_B)$ and the bounds in Eqs.~(\ref{eq:ratio_upperbound1}) and (\ref{eq:ratio_lowerbound2}) for the $d=3$ Dicke model as a representative example. The results exhibit the expected $\sim 1/N$ scaling and agree with the bounds. As shown in the inset, the purity $\Tr[\hat{\rho}_B^2]$ remains below unity for all $N$ (approaching $0.6514$ as $N\to\infty$), confirming the case $\delta_\infty>0$. Figure~\ref{fig:collapse} plots $a^{-1}[1-(\mathcal{E}_B/E_B)]$ versus $N$ for $d=2$--$4$ Dicke and TC models, where $a$ is the constant in Eq.~(\ref{eq:theorem1}) for each model. All data collapse onto a single line, confirming the universal $1/N$ scaling. Numerical calculations were performed using QuTiP \cite{Lambert2026QuTiP5} package.

For the case where the residual population vanishes in the large-$N$ limit, i.e., $\delta_\infty = 0$, Eq.~(\ref{eq:ratio_fastconv}) leads directly to the corollary:

\noindent \textit{Corollary 2}---If $\hat{\rho}_B$ becomes pure in the limit $N \rightarrow \infty$, i.e., $\delta_\infty = 0$, the ratio $\mathcal{E}_B/E_B$ can asymptotically approach unity faster than $\sim 1/N$, as the deviation scales as $\Delta \delta(N)/N$ with $\Delta \delta(N) \to 0$.

\begin{figure}[t!]
\centering
\includegraphics[width=0.7 \columnwidth]{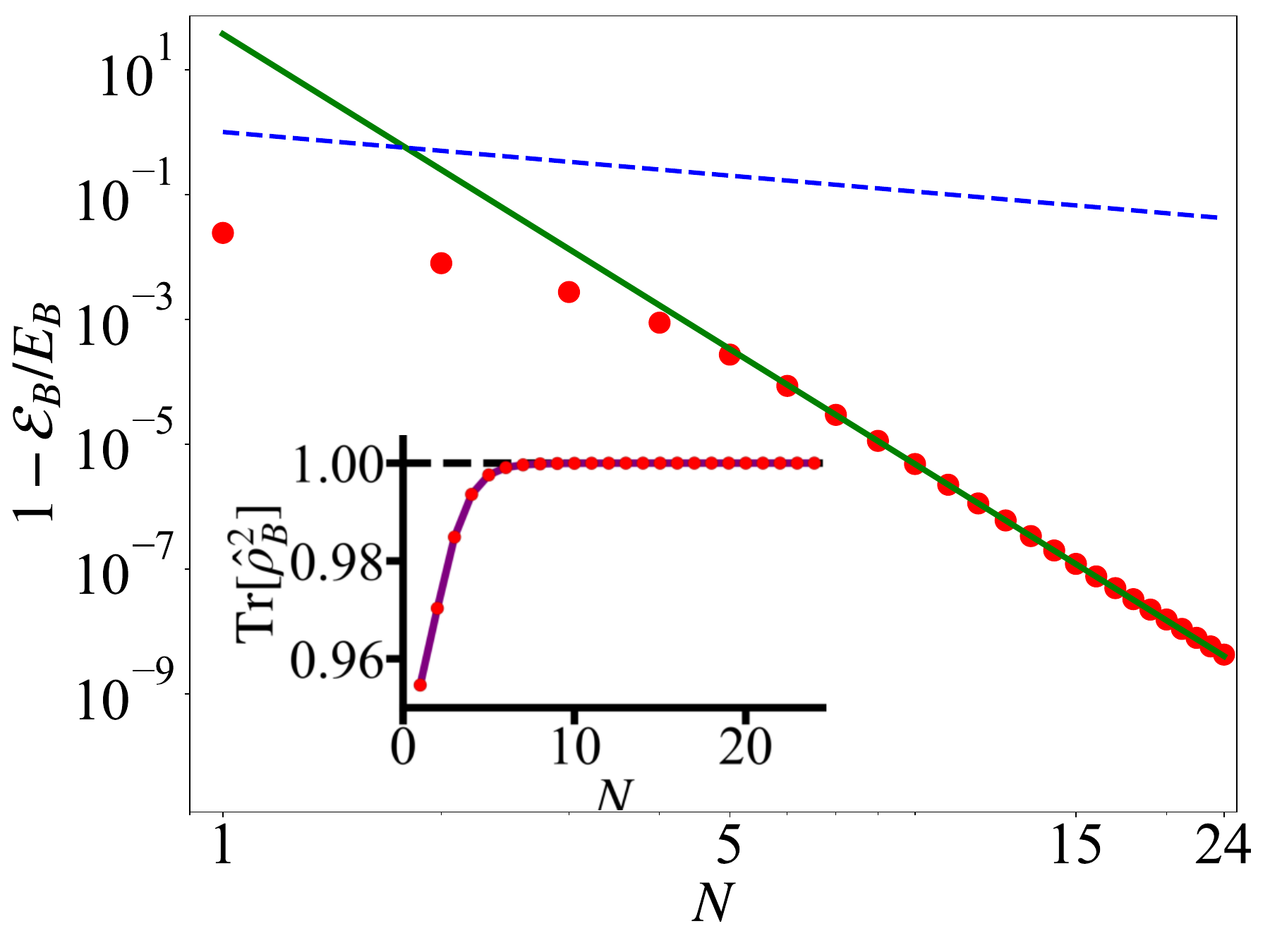}
\caption{Demonstration of fast convergence to asymptotic freedom with power-law scaling $\sim 1/N^{b}$ with $b>1$. The green solid line indicates a fit $\sim 1/N^{7.2}$, while the blue dashed line shows the $\sim 1/N$ scaling for reference. The inset displays the purity as a function of $N$. The parameters are set to $\Omega = \omega_B = 5$ and $\gamma = 1$
}
\label{fig:faster_powerlaw}
\end{figure}

\textit{Example 2: Convergence to asymptotic freedom faster than $1/N$ scaling}---Corollary~2 shows the rate is set by how fast 
$\delta(N) \to 0$, making protocol design a lever 
for faster-than-$1/N$ convergence. We illustrate 
this with open charging of $N$ identical qubits by an engineered bath, where the dynamics of the $N$-qubit system is described by the Gorini–Kossakowski–Sudarshan–Lindblad (GKLS) master equation: $\dot{\hat{\rho}}_B = -i [\hat{H}_B, \hat{\rho}_B] + \frac{\gamma}{2} (2 \hat{L}\hat{\rho}_B\hat{L}^\dagger - \hat{L}^\dagger\hat{L}\hat{\rho}_B - \hat{\rho}_B\hat{L}^\dagger\hat{L})$ where $\gamma$ is the dissipation rate. The battery Hamiltonian as $\hat{H}_B = \omega_B \hat{S}_z + \Omega \hat{S}_x$ (up to an additive constant that sets the ground-state energy to zero), and consider an engineered bath described by the jump operator $\hat{L} = \hat{S}_z - i \hat{S}_y$ with collective spin operators $\hat{S}_{\alpha} \equiv \sum_{i=1}^N \hat{\sigma}_{\alpha,\, i}/2$ ($\alpha = x$, $y$, $z$) defined as a sum of the Pauli matrices acting on the $i$th spin. This choice of $\hat{L}$ drives each qubit toward the coherent superposition state $\propto |0\rangle_i + |1\rangle_i$, which is the steady (dark) state of the dissipative dynamics without $\hat{H}_B$~\cite{Diehl2008,Watanabe2012,Caballar2014}; in the presence of $\hat{H}_B$, the steady state is modified by the competition between the unitary and dissipative dynamics. Starting from $\hat{\rho}_B = \bigotimes_{i=1}^N |0\rangle\langle 0|_i$, we solve the GKLS equation to obtain the steady state and evaluate the corresponding energy $E_B$, ergotropy $\mathcal{E}_B$, and purity. We have verified numerically that the same steady state is reached for a wide range of initial states, including the ground state of $\hat{H}_B$. While $\gamma$ sets the relaxation timescale, the steady state itself is independent of $\gamma$.

\begin{figure}
\centering
\includegraphics[width=0.7 \columnwidth]{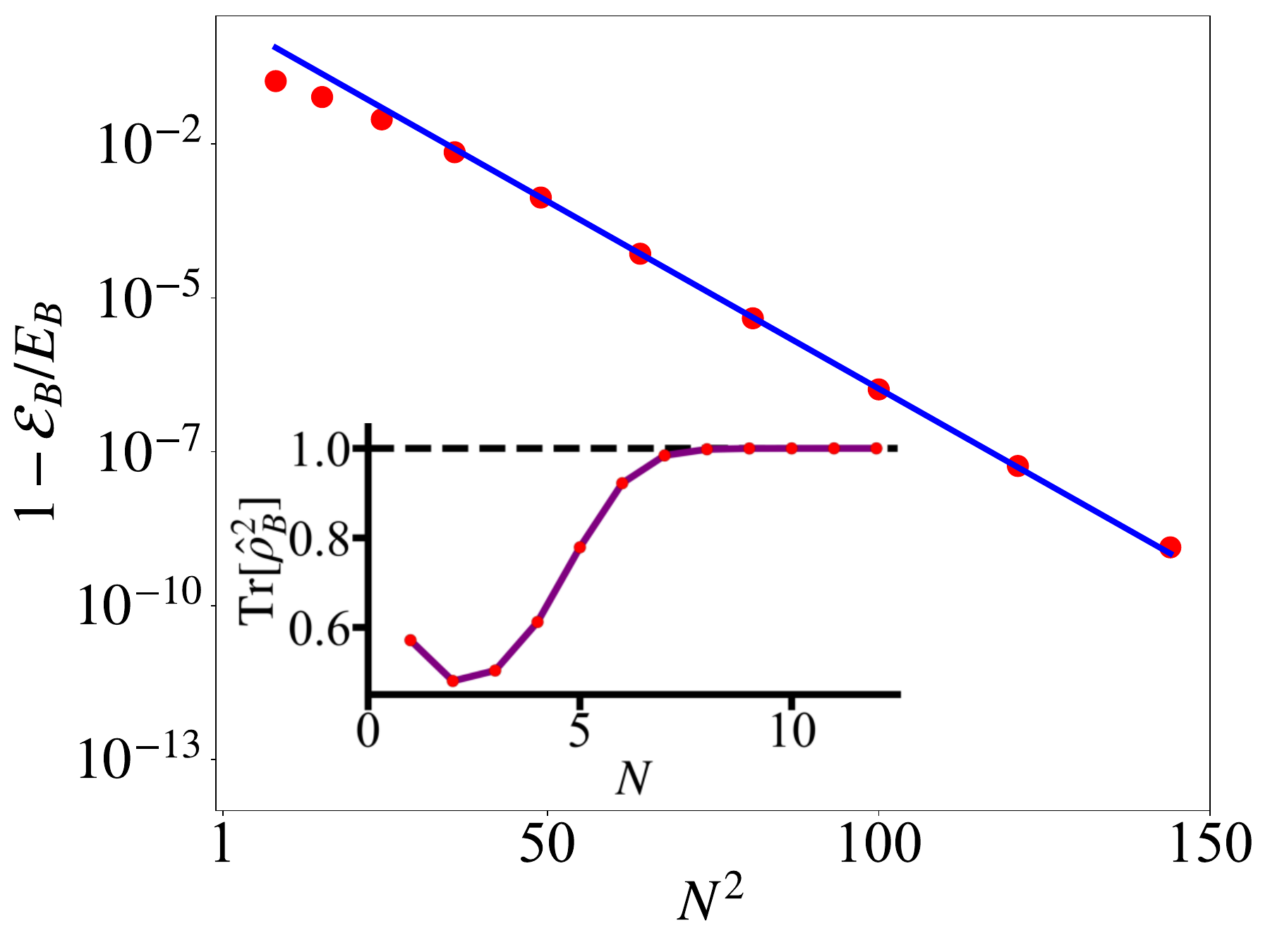}
\caption{Demonstration of fast convergence to asymptotic freedom with exponential scaling. The blue solid line indicates a fit $\sim \exp{(- b N^2)}$ with $b=0.06$. The inset displays the purity as a function of $N$. The parameters are set to $\Omega = 0$, $\omega_B = 10$, and $\gamma = 1$.
}
\label{fig:faster_exp}
\end{figure}

Since the purity $\Tr[\hat{\rho}_B^2]$ approaches unity with increasing $N$ under this charging scheme (see the insets of Figs.~\ref{fig:faster_powerlaw} and \ref{fig:faster_exp}), the convergence to asymptotic freedom is faster than $\sim 1/N$. Figure~\ref{fig:faster_powerlaw} shows that $\mathcal{E}_B/E_B$ approaches unity with a power-law scaling $\sim 1/N^b$ with $b > 1$ when both $\Omega$ and $\omega_B$ are nonzero. Furthermore, Fig.~\ref{fig:faster_exp} demonstrates an even faster exponential convergence, $\sim \exp{(- b N^2)}$ with $b>0$ for $\Omega = 0$ (i.e., in the absence of the transverse field) and $\omega_B \ne 0$.

\textit{Conclusion}---We have proved that the approach to asymptotic freedom for collectively charged quantum batteries has a universal scaling law $\sim 1/N$ when the battery state is mixed in the large-$N$ limit. Furthermore, we have shown this asymptotic behavior can be surpassed by charging protocols in which the battery state approaches a pure state with increasing $N$. This establishes a universal benchmark for the energy efficiency of collective charging and identifies the purity of the reduced battery state as the key figure 
of merit for protocol design. We have also derived rigorous bounds for the ergotropy-to-energy ratio for finite $N$ complementing the asymptotic results. The results we have presented also open up several interesting questions such as --- the protocol we have found for going beyond $\sim 1/N$ scaling required dissipative charging, can this behavior emerge also from collective unitary charging of batteries? Finally, since the vanishing or saturation of 
$\delta_\infty$ separates the generic $1/N$ regime from 
accelerated convergence, it would be valuable to 
determine which physical features of a charging protocol 
control this distinction across the many QB models in 
the literature.

\bigskip
\begin{acknowledgments} 
\emph{Acknowledgements}---This work is supported by NSF of China (Grant No.~12375039) and by the Zhejiang University 100 Plan. B.P.V. acknowledges support from MATRICS Grant No. MTR/2023/000900 from Anusandhan National Research Foundation, Government
of India.
\end{acknowledgments} 

\emph{Data availability}---The data that support the findings of
this article are openly available in the referenced Zenodo dataset~\cite{Chen2026dataset}.

\newpage
\centerline{\large\bfseries End Matter}

\appendix

\section{Extensivity of $E_B$}\label{app:extensivity}

Since there is no interaction between the batteries, the total Hamiltonian $\hat{H}_B$ of the $N$-QB system is simply the sum of the individual battery Hamiltonians:
\begin{equation}
  \hat{H}_B = \sum_{i=1}^N \hat{h}_{B_i}\,,\label{eq:hb}
\end{equation}
where $\hat{h}_{B_i}$ is the Hamiltonian of the $i$th battery, and all batteries are identical, $\hat{h}_B = \hat{h}_{B_i}$ for any $i$. The total density operator $\hat{\rho}_B$ of the $N$-QB system is permutation invariant with respect to the batteries. Consequently, the reduced state of the $i$th battery, $\hat{\rho}_{B_i} \equiv \Tr_{\ne B_i} \hat{\rho}_B$, is the same for all $i$. We denote this common reduced local state by $\hat{\rho}_{\mathrm{loc}}$:
\begin{equation}
  \hat{\rho}_{\mathrm{loc}} = \hat{\rho}_{B_i}\label{eq:rholoc}
\end{equation}
for any $i$.

Using Eqs.~(\ref{eq:hb}) and (\ref{eq:rholoc}), the total energy $E_B$ of the $N$-QB system can be expressed as
\begin{align}
  E_B &\equiv \Tr[\hat{H}_B\, \hat{\rho}_B] = \sum_{i=1}^N \Tr[\hat{h}_{B_i}\, \hat{\rho}_B]
  = \sum_{i=1}^N \Tr_{B_i}[\hat{h}_{B_i}\, \hat{\rho}_{B_i}]\nonumber\\
  &= \sum_{i=1}^N \Tr[\hat{h}_B\, \hat{\rho}_{\mathrm{loc}}] = N\, \Tr[\hat{h}_B\, \hat{\rho}_{\mathrm{loc}}]
  \equiv N\, e_B\,,\label{eq:extensivity}
\end{align}
where $e_B$ is the energy of a single battery,
\begin{equation}
  e_B \equiv \Tr[\hat{h}_B\, \hat{\rho}_{\mathrm{loc}}]\,.
\end{equation}
This shows that, under the current setting of collective charging where $N$ identical copies of a QB are treated identically, the total energy is extensive:
\begin{equation}
  E_B = N e_B \propto N\,.
\end{equation}

\section{Asymptotic bounds when $\delta_\infty$ does not exist}\label{app:no_delta_infty}

In this appendix, we derive asymptotic bounds for $\mathcal{E}_B/E_B$ in the remaining case where $\delta(N)$ does not converge as $N \rightarrow \infty$, i.e., where $\delta_\infty$ does not exist.

\textit{Upper bound of $\mathcal{E}_B/E_B$}\,---\,In this case, Eq.~(\ref{eq:ratio_upperbound1}) together with $\delta(N) \ge \delta_{\mathrm{min}}$ yields the asymptotic upper bound
\begin{equation}
  \frac{\mathcal{E}_B}{E_B} \le 1 - \frac{\Delta \epsilon_0}{e_B} \frac{\delta_{\mathrm{min}}}{N}\,, \quad (N \rightarrow \infty)\,.
\end{equation}

\textit{Lower bound of $\mathcal{E}_B/E_B$}\,---\,Since $\delta(N) < 1$ [$\delta(N)$ cannot equal unity because the passive state always has a nonzero population in the ground state, $\eta_0^\downarrow >0$], Eq.~(\ref{eq:ratio_lowerbound}) implies
\begin{equation}
  \frac{\mathcal{E}_B}{E_B} \ge 1 - (d-1) \frac{\Delta\epsilon_{\mathrm{max}}}{e_B} \frac{\delta(N)}{N}
  > 1 - (d-1) \frac{\Delta\epsilon_{\mathrm{max}}}{e_B} \frac{1}{N}\,.
\end{equation}
Therefore, the asymptotic lower bound reads
\begin{equation}
  \frac{\mathcal{E}_B}{E_B} > 1 - (d-1) \frac{\Delta\epsilon_{\mathrm{max}}}{e_B} \frac{1}{N}\,,
  \quad (N \rightarrow \infty)\,.
\end{equation}

Since both the asymptotic upper and lower bounds scale as $1/N$, Eq.~(\ref{eq:theorem1}) follows, with the coefficient $a$ bounded as $\delta_{\mathrm{min}} \Delta\epsilon_0 / e_B \le a < (d-1)\Delta\epsilon_{\mathrm{max}} / e_B$.

\section{Proof of Lemma 1}\label{app:lemma1}

In this appendix, we provide a proof of the following lemma.

\textit{Lemma 1}---
Consider a system consisting of $N$ copies of a $d$-level system with equal level spacing (a qudit). The entire set of $D_d(N)$ states of the $N$-copy system is completely and exactly exhausted by the states corresponding to the lowest $d$ energy eigenvalues, $\bar{\epsilon}_0 = 0$, $\bar{\epsilon}_1 = \omega_B$, $\cdots$, $\bar{\epsilon}_{d-1} = (d-1)\omega_B$, where $\omega_B$ denotes the level spacing of the qudit.

\textit{Proof}\,---\,For $N$ qudits, as discussed in the main text, the total number of states is given by
\begin{equation}
  D_d(N) = \binom{N+d-1}{d-1} = \frac{(N+d-1)!}{(d-1)!\, N!}\,.\nonumber
\end{equation}
We first evaluate the  difference $\Delta_d(N)$ between the number of states for $N$ copies of a $d$-level system and that for $N$ copies of a $(d-1)$-level system:
\begin{align}
  \Delta_d(N) &\equiv\, D_d(N) - D_{d-1}(N) = \binom{N+d-2}{d-1}\,.
\end{align}

The number of states $N_{d-1}$ associated with the $d$th energy eigenvalue $\bar{\epsilon}_{d-1} = (d-1) \omega_B$ of the $N$-qudit system is equal to the number of ways to distribute $d-1$ excitation quanta among $N$ qudits, which is
\begin{equation}
  N_{d-1} = \binom{N+d-2}{d-1}\,.
\end{equation}
Therefore, we obtain
\begin{equation}
  \Delta_d(N) = N_{d-1}\,.\label{eq:Deltad_n}
\end{equation}

We now prove the lemma by induction.
For $d=1$, the total number of states is $D_1(N) = 1$, corresponding to the unique ground state with eigenvalue $\bar{\epsilon}_0 = 0$. This is consistent with $\Delta_1(N) = \binom{N-1}{0} = 1$, and hence
\begin{equation}
  D_1(N) = \Delta_1(N) = N_0\,.\label{eq:Delta1_n}
\end{equation}
Starting from this base case and iteratively applying the relation $\Delta_d(N) = N_{d-1}$, we conclude that, for any positive integer $d$, the total number of states $D_d(N)$ of $N$-qudit system is completely and exactly exhausted by the states corresponding to the lowest $d$ energy eigenvalues, $\bar{\epsilon}_0 = 0$, $\bar{\epsilon}_1 = \omega_B$, $\cdots$, $\bar{\epsilon}_{d-1} = (d-1)\omega_B$. This completes the proof.

\section{Proof of Eq.~(\ref{eq:energysum1})}\label{app:energyspectrum}

\begin{figure}[t]
\centering
\includegraphics[width=0.6 \columnwidth]{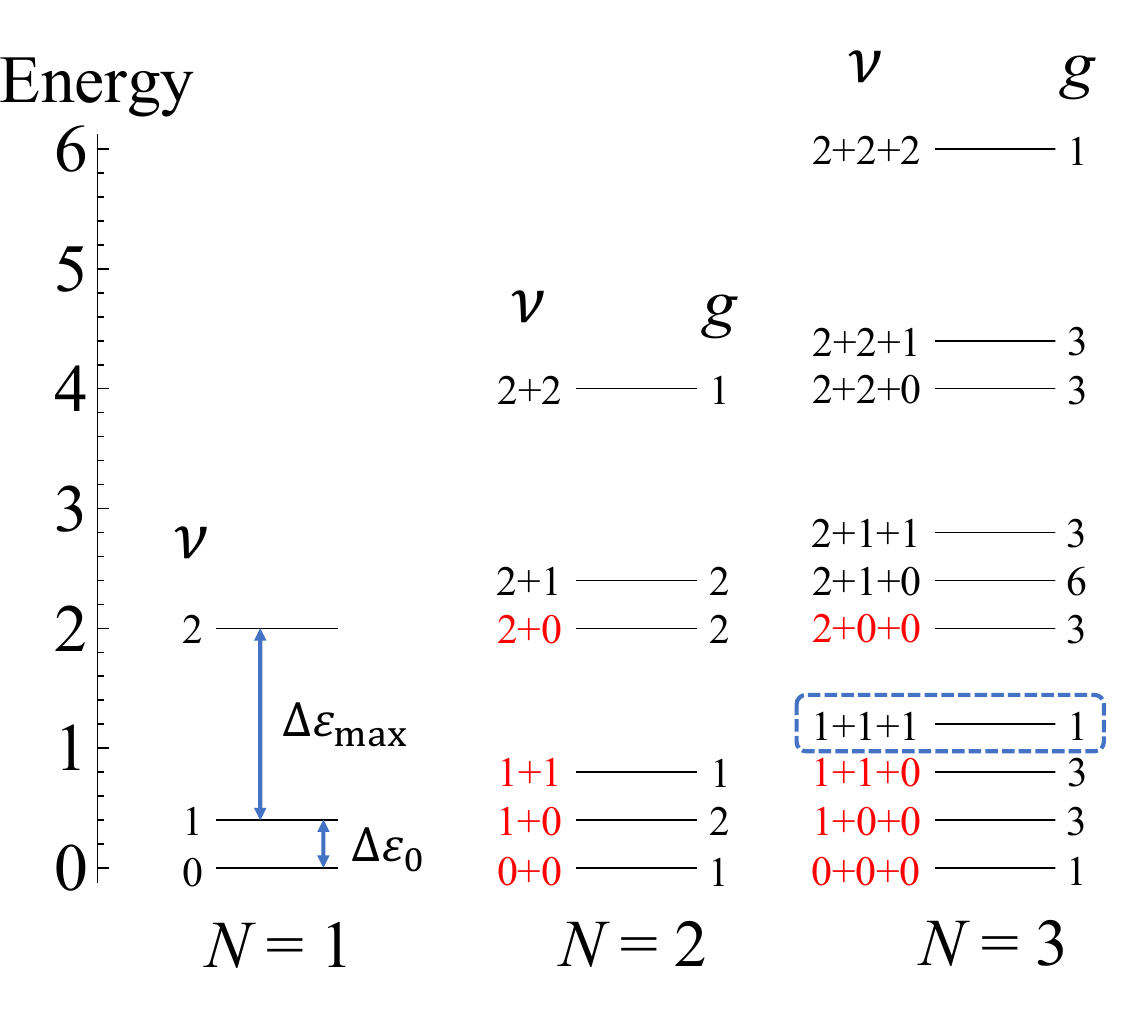}
\caption{Example of energy levels for $N$ identical copies of a $d$-level battery with uneven level spacings ($d=3$). Here, $\nu$ and $g$ denote the excitation number and multiplicity of each level. The values of $\nu \le d-1 =2$ are highlighted in red, whose total multiplicity equals $D_{3}(N)=(N+2)(N+1)/2$. For uneven level spacings, some states with $\nu > d-1$ can have lower energy than some of those with $\nu\le d-1$, as indicated by the blue dashed square.}
\label{fig:energylevels}
\end{figure}

In this appendix, we show Eq.~(\ref{eq:energysum1}) for general QBs with uneven level spacings.

According to Lemma 1, for $d$-level system with equal level spacing, the entire set of $D_d(N)$ states of the $N$-copy system is completely and exactly exhausted by the states corresponding to the lowest $d$ energy eigenvalues whose excitation numbers $\nu$ are $\nu=0$, $1$, $\cdots$, $d-1$. In the case of general QBs with uneven level spacings, on the other hand, there is a possibility that some states with $\nu > d-1$ have lower energy than some of those with $\nu \le d-1$ as indicated by the blue dashed square in Fig.~\ref{fig:energylevels}. Therefore, the total energy of the lowest $D_d(N)$ states is upper bounded by the total energy of the $D_d(N)$ states with $\nu \le d-1$:
\begin{equation}
  \sum_{i=0}^{D_d - 1} \epsilon_i^\uparrow \le \mbox{(total energy of states with $\nu \le d-1$)}\,,\nonumber
\end{equation}
as claimed by Eq.~(\ref{eq:energysum1}).


\clearpage

\setcounter{section}{0}
\setcounter{subsection}{0}
\setcounter{equation}{0}
\setcounter{figure}{0}
\setcounter{table}{0}

\renewcommand{\thesection}{\Roman{section}}
\renewcommand{\thesubsection}{\Alph{subsection}}
\renewcommand{\theequation}{S\arabic{equation}}
\renewcommand{\thefigure}{S\arabic{figure}}
\renewcommand{\thetable}{S\arabic{table}}

\setcounter{section}{0}
\onecolumngrid

\begin{center}
\begin{center}
{\Large\bfseries Supplemental Material for ``Scaling law of asymptotic freedom in collective charging of quantum batteries''}
\end{center}

\vspace{0.5cm}

Gentaro Watanabe$^{1,2}$, Chunlin Chen$^{1}$, and B. Prasanna Venkatesh$^{3}$

\smallskip
{\small \textit{
$^{1}$ Department of Physics and Modern Mechanics, Zhejiang University, Hangzhou, Zhejiang 310027, China\\
$^{2}$ Zhejiang Province Key Laboratory of Quantum Technology and Device, Zhejiang University, Hangzhou, Zhejiang 310027, China\\
$^{3}$ Indian Institute of Technology Gandhinagar, Palaj, Gujarat 382055, India}
}

\vspace{0.5cm}
\end{center}

\noindent\textbf{Abstract.}
This supplemental material provides further numerical evidence for the fast convergence to asymptotic freedom with exponential scaling in $N^2$ in the open charging scheme studied in the main text.

\vspace{0.5cm}

\twocolumngrid


\section*{I. Further demonstration of exponential scaling in $N^2$}

In the main text, we demonstrated that the ergotropy-to-energy ratio $\mathcal{E}_B/E_B$ asymptotically approaches unity as $\sim \exp{(-b N^2)}$ with $b>0$ for the open charging of $N$ identical qubits by an engineered bath. In that setup, the battery Hamiltonian is given by
\begin{equation}
  \hat{H}_B = \omega_B \left(\hat{S}_z + \frac{N}{2}\right)\,,
\end{equation}
and the dynamics is governed by the GKLS master equation $\dot{\hat{\rho}}_B = -i [\hat{H}_B, \hat{\rho}_B] + \frac{\gamma}{2} (2 \hat{L}\hat{\rho}_B\hat{L}^\dagger - \hat{L}^\dagger\hat{L}\hat{\rho}_B - \hat{\rho}_B\hat{L}^\dagger\hat{L})$, with $\hat{L} = \hat{S}_z - i \hat{S}_y$, where $\hat{S}_\alpha \equiv \sum_{i=1}^N \hat{\sigma}_{\alpha, i}/2$ ($\alpha = x$, $y$, $z$) denotes the collective spin operator, as defined in the main text. From the steady state obtained by solving the GKLS master equation, we evaluate the energy $E_B$, ergotropy $\mathcal{E}_B$, and purity $\Tr[\hat{\rho}_B^2]$ of the battery state $\hat{\rho}_B$. We set the dissipation rate $\gamma = 1$ as in the main text.

\begin{figure}[H]
\centering
\includegraphics[width=0.82\columnwidth]{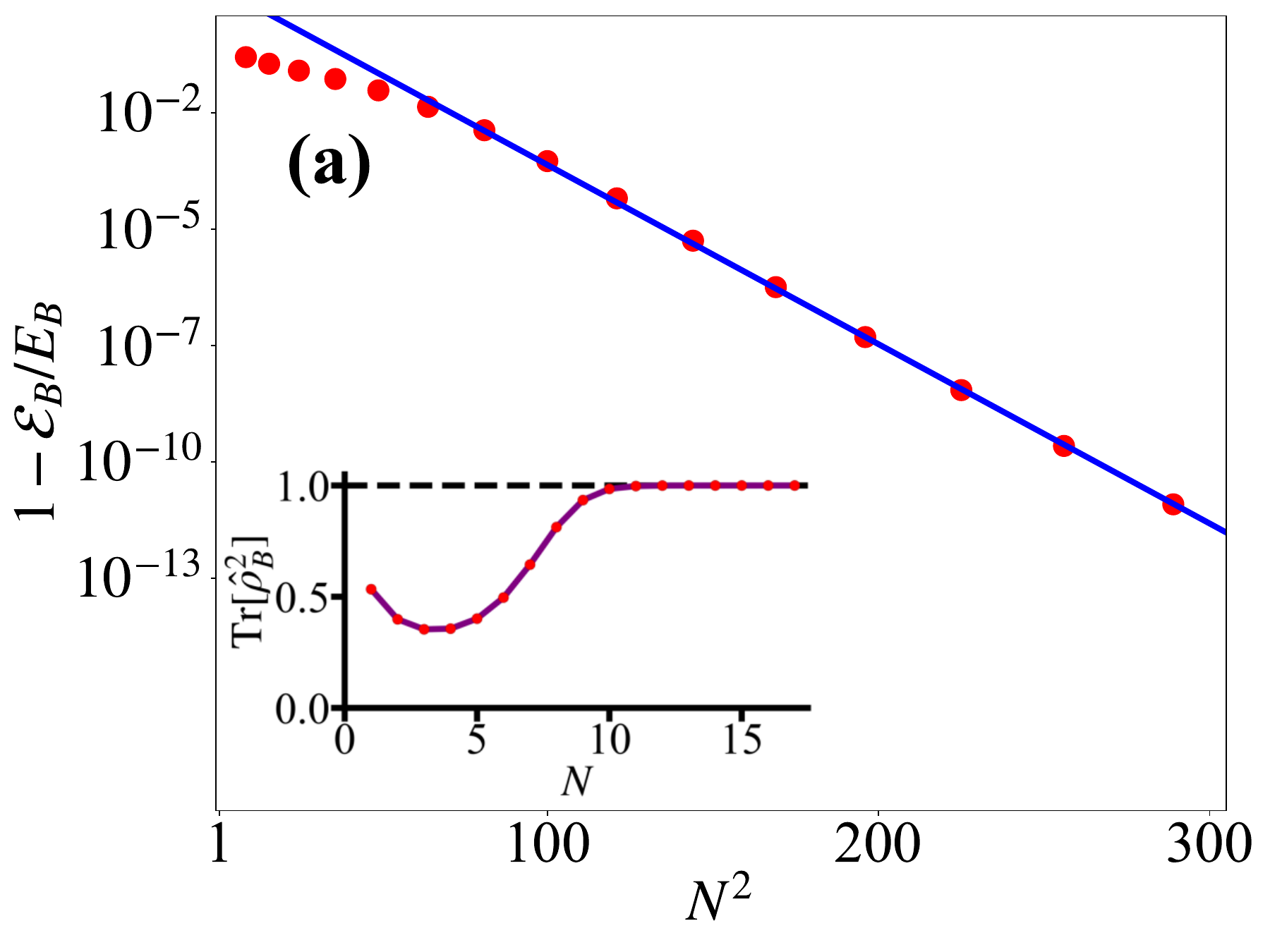}\\[2pt]
\includegraphics[width=0.82\columnwidth]{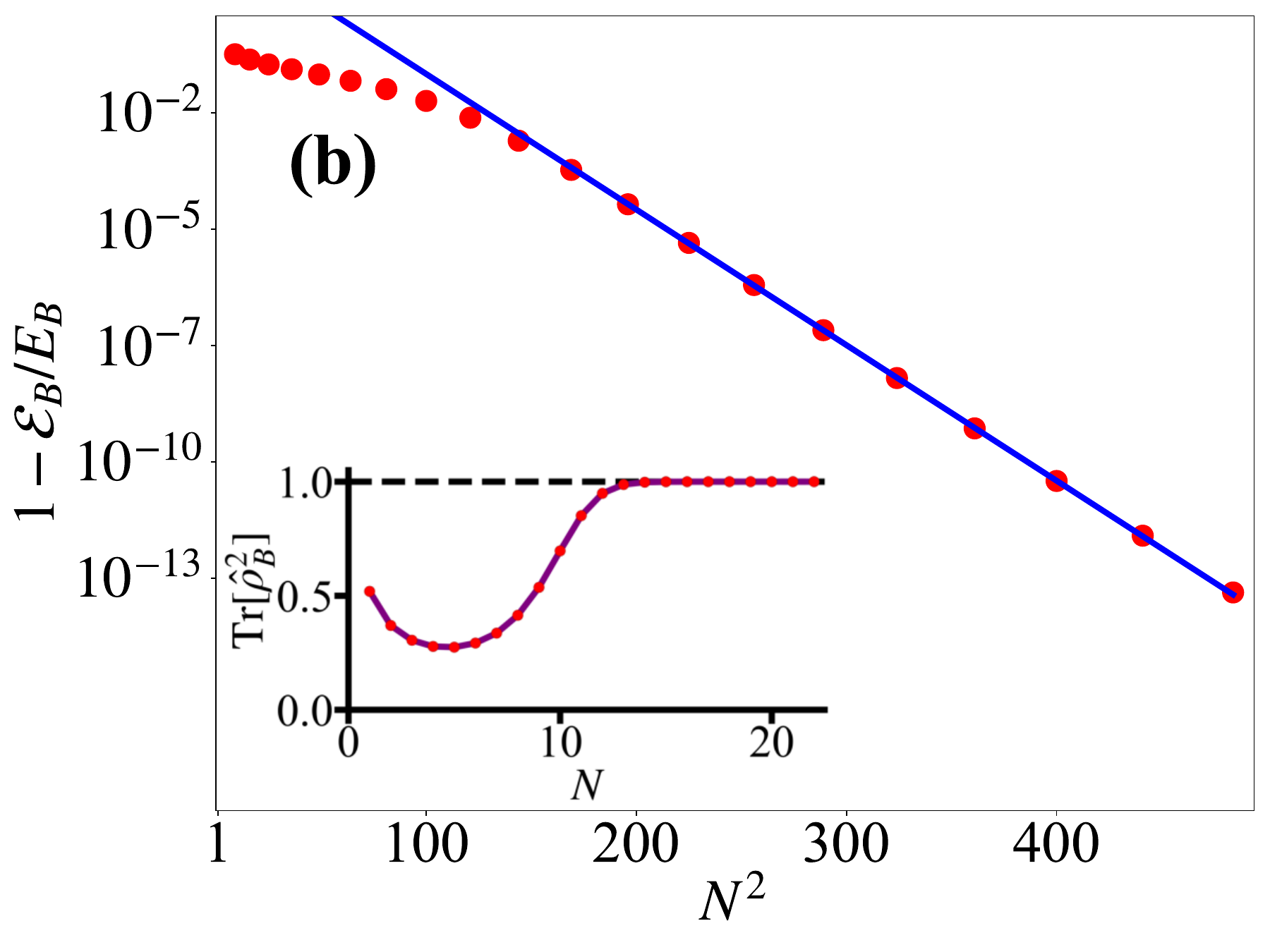}\\[2pt]
\caption{Additional demonstration of fast convergence to asymptotic freedom with exponential scaling in $N^2$. The blue solid lines represent the asymptotic scaling form $\sim \exp{(- b N^2)}$ with $b = 0.04$ [panel (a)] and $0.03$ [panel (b)]. The insets show the purity $\Tr[\hat{\rho}_B^2]$ as a function of $N$. The parameters are set to $\omega_B = 15$ [panel (a)] and $20$ [panel (b)], with $\gamma = 1$ in both cases.
}
\label{fig:S1}
\end{figure}

Figure~\ref{fig:S1} shows the $N$-dependence of $1 - \mathcal{E}_B/E_B$ for $\omega_B = 15$ and $20$. As in the case of $\omega_B = 10$ presented in the main text, the purity $\Tr[\hat{\rho}_B^2]$ rapidly approaches unity with increasing $N$, and $1 - \mathcal{E}_B/E_B$ is already well described by the asymptotic scaling form $\sim \exp{(-b N^2)}$ represented by the solid blue lines even for relatively small $N \gtrsim 10$.

\end{document}